\documentclass{aa}
\usepackage{txfonts}
\usepackage{lscape}
\bibpunct{(}{)}{;}{a}{}{,}    
\usepackage{graphicx}
\usepackage{adjustbox}
\usepackage[breaklinks=true, colorlinks=true, urlcolor=blue, citecolor=blue, linkcolor=blue]{hyperref}

\begin{document}

\title{Characterizing chemical abundance ratios in extremely metal-poor star-forming galaxies in DESI EDR}

\author{
        I.~A.~Zinchenko\inst{\ref{LMU},\ref{MAO}} \and
        M.~Sobolenko\inst{\ref{MAO}} \and
        J.~M.~V\'{i}lchez\inst{\ref{IAA}} \and
        C.~Kehrig\inst{\ref{IAA}}
}
       
\institute{
Faculty of Physics, Ludwig-Maximilians-Universit\"{a}t, Scheinerstr. 1, 81679 Munich, Germany \label{LMU}
\and
Main Astronomical Observatory, National Academy of Sciences of Ukraine, 
27 Akademika Zabolotnoho St., 03143, Kyiv, Ukraine\label{MAO}
\and
Instituto de Astrof\'{i}sica de Andaluc\'{i}a (CSIC), Apartado 3004, 18080 Granada, Spain \label{IAA} 
\label{LU}
}

\abstract{%
We present a search for galaxies in the local Universe with extremely low oxygen abundance, that is, more than 25 times lower than solar, which corresponds to 12 + $\log$(O/H) < 7.3. To determine the oxygen abundance, we apply the direct T$_e$ method for objects where the [\ion{O}{III}]$\lambda$4363 line is detected. 
We identified 21 extremely metal-poor galaxies in the early data release of the Dark Energy Spectroscopic Instrument (DESI EDR), for some of which we also derived N/O, Ne/O, Ar/O, and S/O ratios. We find that many DESI galaxies with extremely low oxygen abundance exhibit a higher N/O ratio in comparison to the reference low-metallicity sample collected from the literature. We suggest that the elevation in N/O ratio may be explained by a contamination with metal-rich gas caused by gas inflow or a merger event. Moreover, contrary to some recent studies, we find that Ar/O and S/O ratios are enhanced as well, while the Ne/O ratio does not show such elevation. 
One of the galaxies, J0713+5608, has a remarkably low oxygen abundance of 6.978$\pm$0.095~dex. This measurement aligns with the lowest known oxygen abundances in galaxies to date. Given the relatively high uncertainty, this galaxy may have the lowest oxygen abundance ever found. Additionally, J0713+5608 exhibited an enhanced N/O ratio compared to the typical N/O ratio observed in metal-poor galaxies within the local Universe.
}


\titlerunning{Extremely low-metallicity star-forming galaxies in the DESI EDR}
\authorrunning{Zinchenko et~al.}
\maketitle


\section{Introduction}

Extremely metal-poor galaxies are relatively uncommon objects in the local Universe, yet they serve as excellent laboratories for the examination of physical conditions and processes in the primeval Universe and during the epoch of reionization. Consequently, investigating these galaxies holds significant importance for advancing our understanding of the early stages of galaxy evolution.

Several such galaxies have been identified in the local Universe through various studies. Since the confirmation of an extremely low oxygen abundance in the blue compact dwarf (BCD) galaxy IZw18, namely of about 12+$\log$(O/H) = 7.11 \citep[see, e.g.,][]{Izotov1998,Kehrig2016}, efforts to discover galaxies with even lower metallicity have been undertaken. For instance, \citet{Skillman2013} measured the oxygen abundance in the \ion{H}{II}~region of the dwarf galaxy Leo~P, finding 7.17$\pm$0.04, while \citet{Hsyu2017} derived an oxygen abundance of 7.13$\pm$0.08~dex for the BCD galaxy Little Cub. The BCD galaxy AGC198691 was found to possess a lower oxygen abundance of 7.02$\pm$0.03~dex by \citet{Hirschauer2016}, or 7.06$\pm$0.03~dex in a more recent study by \citet{Aver2022}. Similar oxygen abundances (12 + $\log$(O/H) = 7.035$\pm$0.026) were observed in J1234+3901 by \citet{Izotov2019}, in J2229+2725 (12+$\log$(O/H) = 7.085$\pm$0.031) by \citet{Izotov2021}, and in J2229+2725 (12+$\log$(O/H) = 7.085$\pm$0.031) by \citet{Izotov2024}.

To date, several galaxies have been reported with oxygen abundances slightly below 12+$\log$(O/H)~=~7.0. \citet{Annibali2019} determined a value of 6.96$\pm$0.09dex in one of the \ion{H}{ii}~regions in the dwarf irregular galaxy DDO~68. J0811+4730, with 12+$\log$(O/H) = 6.98$\pm$0.02, was identified by \citet{Izotov2018}. \citet{Kojima2020} reported an even lower oxygen abundance of 6.90$\pm$0.03dex for HSC~J1631+4426. However, \citet{Thuan2022} reobserved this galaxy with the Large Binocular Telescope (LBT) and determined a significantly higher metallicity of 12+$\log$(O/H) = 7.14$\pm$0.03.

In the present work, we aim to search for extremely metal-poor galaxies (12 + $\log$(O/H) < 7.3) in the local Universe ($z < 0.5$) in the early data release of the Dark Energy Spectroscopic Instrument~\citep[DESI EDR;][]{DESI2023} using the direct T$_e$ method and explore abundances of O, N, Ne, Ar, S, and Fe, the emission lines of which are available in the optical spectrum.
The paper is organized as follows. In Sect.~\ref{sect:data} we provide a detailed description of the data used for further analysis. Section~\ref{sect:abundance} describes the method of determination of chemical abundances and other physical properties. In Sect.~\ref{sect:analysis} we discuss the obtained sample of extremely metal-poor galaxies. In Sect.~\ref{sect:J0713+5608} we describe the galaxy with the lowest oxygen abundance. Finally, Sect.~\ref{section:Summary} presents a summary of our main findings.

\begin{table*}
\caption{General characteristics of our sample of galaxies from the DESI EDR}     
\label{table:info}      
\centering                          
\begin{tabular}{l c c c c c c c r r r r}        
\hline\hline                 
Name & Target ID & RA & DEC & Redshift & mag$_g$ & mag$_r$ & M$_g$ & Exptime \\    
     &           &deg & deg &          & mag     & mag     & mag   & s \\ 
\hline                        
 J1143-0139 & 39627745398883762 & 175.815730 & -1.651681 & 0.1445 & 20.65 &20.45 &-18.31 &  4215 \\ 
 J0941+3209 & 39628526860634476 & 145.296859 & 32.156355 & 0.0213 & 19.79 &19.67 &-15.01 &  5623 \\ 
 J1333+3326 & 39632951142518497 & 203.301564 & 33.438166 & 0.1253 & 23.05 &23.02 &-15.60 &   617 \\ 
 J0224-0328 & 39627700792463174 &  36.141673 & -3.474988 & 0.1770 & 23.89 &23.75 &-15.51 & 10800 \\ 
 J1212-0044 & 39627769679712162 & 183.167657 & -0.748498 & 0.0404 & 23.17 &23.84 &-13.03 & 10581 \\ 
 J1003+0123 & 39627823496826637 & 150.926393 &  1.385053 & 0.0998 & 23.68 &23.80 &-14.48 & 13500 \\ 
 J1225+3151 & 39628517175986127 & 186.274146 & 31.862207 & 0.0443 & 23.35 &23.55 &-13.04 &  2991 \\ 
 J0847+3257 & 39632940019221902 & 131.901191 & 32.953882 & 0.1418 & 23.53 &23.65 &-15.39 &  4555 \\ 
 J0901+3336 & 39632950190410845 & 135.403924 & 33.603166 & 0.1162 & 22.05 &22.06 &-16.44 &  5508 \\ 
 J0651+3843 & 39633052476903702 & 102.917068 & 38.733027 & 0.0662 & 23.23 &23.42 &-14.03 &  9184 \\ 
 J0713+5608 & 39633338331299912 & 108.360899 & 56.142043 & 0.0389 & 22.63 &22.51 &-13.48 & 13500 \\ 
 J0923+6451 & 39633440282250119 & 140.880315 & 64.853146 & 0.0054 & 20.71 &20.48 &-11.13 &  9000 \\ 
 J1505+3146 & 39628517750604871 & 226.399556 & 31.777634 & 0.0543 & 20.51 &20.57 &-16.32 &  2066 \\ 
 J1802+6439 & 39633441217579195 & 270.597146 & 64.656143 & 0.0494 & 19.20 &19.00 &-17.43 &   285 \\ 
 J1434-0055 & 39627764235503713 & 218.584632 & -0.923267 & 0.1370 & 22.74 &22.44 &-16.11 &  1566 \\ 
 J1001+0241 & 39627853687427435 & 150.439479 &  2.691775 & 0.2192 & 23.10 &22.34 &-16.77 &  2881 \\ 
 J1257+2348 & 39628346086134793 & 194.450750 & 23.810648 & 0.0887 & 23.17 &23.12 &-14.73 &  2653 \\ 
 J1256+2433 & 39628362649440128 & 194.170086 & 24.551167 & 0.0488 & 21.65 &21.83 &-14.96 &   870 \\ 
 J1301+2505 & 39628373655292169 & 195.266315 & 25.099504 & 0.0257 & 20.75 &20.71 &-14.46 &   870 \\ 
 J1651+3356 & 39632961900904930 & 252.865044 & 33.936124 & 0.1127 & 23.88 &25.26 &-14.54 &   842 \\ 
 J1536+4346 & 39633145200381558 & 234.133969 & 43.767690 & 0.4662 & 23.80 &23.82 &-17.70 &  1499 \\ 
\hline                                   
\end{tabular}
\end{table*}

\begin{table*}
\caption{Electron densities, [\ion{O}{III}] temperatures t$_3$, and element abundances in the DESI EDR galaxies}    
\label{table:abundance}      
\centering                          
\begin{adjustbox}{width=\linewidth}
\begin{tabular}{l c c c c c c c c c}        
\hline\hline                 
 Target ID  & n$_e$([\ion{S}{II}]) &  n$_e$([\ion{O}{II}])  &  t3           & 12+$\log$(O/H) & 12+$\log$(N/H) & 12+$\log$(Ne/H) & 12+$\log$(Ar/H) & 12+$\log$(S/H) \\  
            & cm$^{-3}$   &  cm$^{-3}$  &10$^4$K        & dex            & dex            & dex             & dex             & dex            \\
\hline                        
 J1143-0139 & 265$\pm$212 & 123$\pm$112 & 2.24$\pm$0.14 & 7.219$\pm$0.041 & 5.784$\pm$0.093 & 6.459$\pm$0.066 & 4.596$\pm$0.077 &     ---         \\
 J0941+3209 &   <100      &   <100      & 2.63$\pm$0.55 & 7.145$\pm$0.095 & 5.830$\pm$0.121 & 6.282$\pm$0.163 &   ---           & 5.617$\pm$0.070 \\
 J1333+3326 &     ---     &   <100      & 2.78$\pm$0.48 & 7.250$\pm$0.093 &   ---           &   ---           &   ---           &     ---         \\
 J0224-0328 &     ---     &   <100    & 2.13$\pm$0.30 & 7.286$\pm$0.096 &   ---           & 6.607$\pm$0.140 &   ---           &     ---         \\
 J1212-0044 &   <100      &  79$\pm$112 & 2.22$\pm$0.35 & 7.235$\pm$0.088 & 5.679$\pm$0.150 & 6.434$\pm$0.142 & 4.923$\pm$0.119 & 5.640$\pm$0.079 \\
 J1003+0123 &     ---     & 205$\pm$172 & 2.96$\pm$0.57 & 7.111$\pm$0.079 &   ---           & 6.295$\pm$0.127 &   ---           &     ---         \\
 J1225+3151 &   <100      &    <100     & 2.31$\pm$0.30 & 7.260$\pm$0.085 &   ---           & 6.592$\pm$0.113 & 5.041$\pm$0.134 & 5.920$\pm$0.080 \\
 J0847+3257 &     ---     & 140$\pm$237 & 2.30$\pm$0.30 & 7.268$\pm$0.086 &   ---           & 6.528$\pm$0.130 &   ---           &     ---         \\
 J0901+3336 &   <100      &    <100     & 2.29$\pm$0.15 & 7.290$\pm$0.038 &   ---           & 6.555$\pm$0.070 &   ---           &     ---         \\
 J0651+3843 & 217$\pm$237 &   <100      & 2.44$\pm$0.43 & 7.232$\pm$0.095 & 6.079$\pm$0.153 & 6.488$\pm$0.132 &   ---           & 5.822$\pm$0.122 \\
 J0713+5608 & 124$\pm$234 &  70$\pm$65  & 2.73$\pm$0.48 & 6.978$\pm$0.095 & 5.864$\pm$0.131 & 6.125$\pm$0.155 & 4.603$\pm$0.161 &     ---         \\
 J0923+6451 &  42$\pm$60  &  18$\pm$17  & 2.06$\pm$0.13 & 7.170$\pm$0.051 & 5.898$\pm$0.071 & 6.471$\pm$0.071 & 4.899$\pm$0.061 & 5.755$\pm$0.037 \\
 J1505+3146 &   <100      & 272$\pm$212 & 2.37$\pm$0.10 & 7.139$\pm$0.029 &   ---           & 6.450$\pm$0.044 & 4.464$\pm$0.097 & 5.397$\pm$0.033 \\
 J1802+6439 &  70$\pm$112 & 172$\pm$112 & 2.80$\pm$0.47 & 7.208$\pm$0.071 & 6.022$\pm$0.104 & 6.526$\pm$0.123 &   ---           &     ---         \\ 
 J1434-0055 &     ---     & 165$\pm$212 & 2.53$\pm$0.34 & 7.249$\pm$0.084 &   ---           & 6.525$\pm$0.104 &   ---           &     ---         \\
 J1001+0241 &1248$\pm$1424&    <100     & 2.34$\pm$0.12 & 7.168$\pm$0.038 &   ---           & 6.486$\pm$0.048 & 4.934$\pm$0.084 &     ---         \\
 J1257+2348 &     ---     & 117$\pm$112 & 2.89$\pm$0.47 & 7.229$\pm$0.076 &   ---           & 6.526$\pm$0.113 &   ---           &     ---         \\
 J1256+2433 &  37$\pm$65  &   <100      & 2.35$\pm$0.16 & 7.281$\pm$0.044 & 5.916$\pm$0.101 & 6.547$\pm$0.060 & 5.094$\pm$0.078 & 5.772$\pm$0.037 \\
 J1301+2505 &   <100      &  28$\pm$34  & 2.22$\pm$0.31 & 7.116$\pm$0.078 &   ---           & 6.211$\pm$0.127 & 4.786$\pm$0.119 & 5.482$\pm$0.062 \\
 J1651+3356 &     ---     & 221$\pm$237 & 2.44$\pm$0.42 & 7.288$\pm$0.099 &   ---           & 6.670$\pm$0.123 &   ---           &     ---         \\
 J1536+4346 &     ---     & 392$\pm$399 & 2.53$\pm$0.36 & 7.238$\pm$0.081 &   ---           & 6.303$\pm$0.134 &   ---           &     ---         \\
\hline                                   
\end{tabular}
\end{adjustbox}
\end{table*}

\begin{table*}
\caption{Ion abundances in the DESI EDR galaxies}             
\label{table:ions}      
\centering                          
\begin{tabular}{l c c c c c c c c c c}        
\hline\hline                 
 Name       & O$^+$/H         & O$^{2+}$/H      &  N$^+$/H        & Ne$^{2+}$/H     & Ar$^{2+}$/H     & S$^+$/H         &  S$^{2+}$/H & \\    
\hline                        
 J1143-0139 & 6.570$\pm$0.060 & 7.109$\pm$0.051 & 5.146$\pm$0.093 & 6.420$\pm$0.066 & 4.543$\pm$0.077 & 4.870$\pm$0.043 &   ---           \\
 J0941+3209 & 6.836$\pm$0.143 & 6.851$\pm$0.115 & 5.519$\pm$0.121 & 6.201$\pm$0.163 &   ---           & 5.266$\pm$0.092 & 5.443$\pm$0.096 \\
 J1333+3326 & 6.422$\pm$0.151 & 7.181$\pm$0.104 &   ---           &   ---           &   ---           &   ---           &   ---           \\
 J0224-0328 & 6.649$\pm$0.128 & 7.172$\pm$0.115 &   ---           & 6.566$\pm$0.140 &   ---           &   ---           &   ---           \\
 J1212-0044 & 6.739$\pm$0.133 & 7.067$\pm$0.109 & 5.183$\pm$0.150 & 6.380$\pm$0.142 & 4.889$\pm$0.119 & 5.072$\pm$0.082 & 5.472$\pm$0.103 \\
 J1003+0123 & 6.638$\pm$0.130 & 6.933$\pm$0.095 &   ---           & 6.237$\pm$0.127 &   ---           &   ---           &   ---           \\
 J1225+3151 & 6.349$\pm$0.115 & 7.203$\pm$0.095 &   ---           & 6.570$\pm$0.113 & 4.933$\pm$0.134 & 4.789$\pm$0.097 & 5.593$\pm$0.091 \\
 J0847+3257 & 6.307$\pm$0.128 & 7.218$\pm$0.094 &   ---           & 6.508$\pm$0.130 &   ---           &   ---           &   ---           \\
 J0901+3336 & 6.601$\pm$0.059 & 7.190$\pm$0.045 &   ---           & 6.519$\pm$0.070 &   ---           & 4.926$\pm$0.057 &   ---           \\
 J0651+3843 & 6.401$\pm$0.135 & 7.162$\pm$0.107 & 5.275$\pm$0.153 & 6.462$\pm$0.132 &   ---           & 4.883$\pm$0.105 & 5.523$\pm$0.145 \\
 J0713+5608 & 6.468$\pm$0.147 & 6.817$\pm$0.115 & 5.356$\pm$0.131 & 6.072$\pm$0.155 & 4.567$\pm$0.161 & 4.960$\pm$0.091 &   ---           \\
 J0923+6451 & 6.511$\pm$0.071 & 7.063$\pm$0.061 & 5.251$\pm$0.071 & 6.432$\pm$0.071 & 4.846$\pm$0.061 & 5.020$\pm$0.045 & 5.535$\pm$0.046 \\
 J1505+3146 & 6.300$\pm$0.044 & 7.071$\pm$0.033 &   ---           & 6.424$\pm$0.044 & 4.374$\pm$0.097 & 4.646$\pm$0.032 & 5.033$\pm$0.045 \\
 J1802+6439 & 6.819$\pm$0.112 & 6.981$\pm$0.086 & 5.629$\pm$0.104 & 6.455$\pm$0.123 &   ---           & 5.378$\pm$0.081 &   ---           \\ 
 J1434-0055 & 6.404$\pm$0.117 & 7.182$\pm$0.094 &   ---           & 6.499$\pm$0.104 &   ---           &   ---           &   ---           \\
 J1001+0241 & 5.809$\pm$0.069 & 7.148$\pm$0.039 &   ---           & 6.477$\pm$0.048 & 4.630$\pm$0.084 & 4.316$\pm$0.087 &   ---           \\
 J1257+2348 & 6.554$\pm$0.127 & 7.126$\pm$0.089 &   ---           & 6.489$\pm$0.113 &   ---           &   ---           &   ---           \\
 J1256+2433 & 6.547$\pm$0.057 & 7.192$\pm$0.052 & 5.201$\pm$0.101 & 6.514$\pm$0.060 & 5.028$\pm$0.078 & 5.064$\pm$0.044 & 5.483$\pm$0.048 \\
 J1301+2505 & 6.745$\pm$0.114 & 6.875$\pm$0.099 &   ---           & 6.138$\pm$0.127 & 4.755$\pm$0.119 & 4.993$\pm$0.076 & 5.350$\pm$0.081 \\
 J1651+3356 & 6.269$\pm$0.141 & 7.245$\pm$0.108 &   ---           & 6.653$\pm$0.123 &   ---           &   ---           &   ---           \\
 J1536+4346 & 6.336$\pm$0.129 & 7.180$\pm$0.089 &   ---           & 6.280$\pm$0.134 &   ---           &   ---           &   ---           \\
\hline                                   
\end{tabular}
\end{table*}

\begin{figure*}
\centering
\begin{minipage}{1.00\textwidth}
    \includegraphics[width=0.19\linewidth]{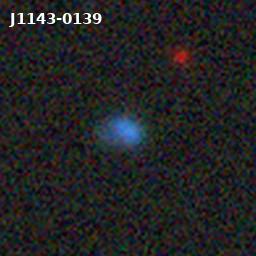}
    \includegraphics[width=0.19\linewidth]{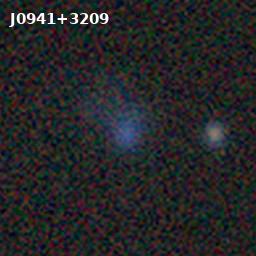}
    \includegraphics[width=0.19\linewidth]{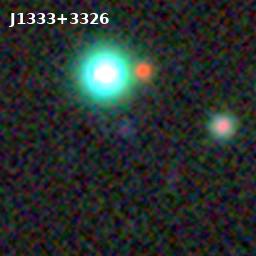}
    \includegraphics[width=0.19\linewidth]{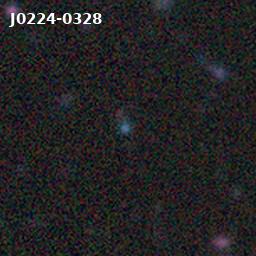}
    \includegraphics[width=0.19\linewidth]{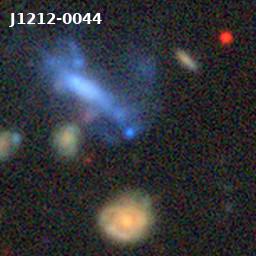}
\end{minipage}
\begin{minipage}{1.00\textwidth}
    \includegraphics[width=0.19\linewidth]{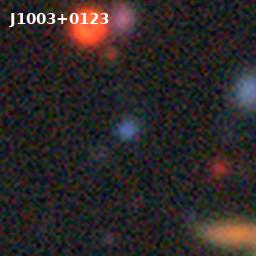}
    \includegraphics[width=0.19\linewidth]{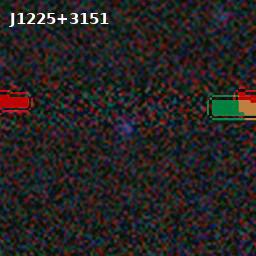}
    \includegraphics[width=0.19\linewidth]{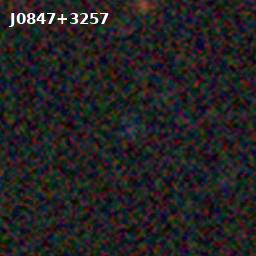}
    \includegraphics[width=0.19\linewidth]{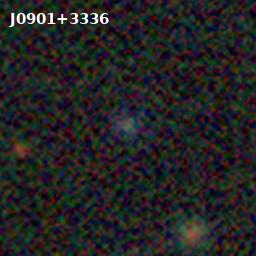}
    \includegraphics[width=0.19\linewidth]{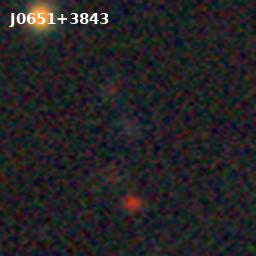}
\end{minipage}
\begin{minipage}{1.00\textwidth}
    \includegraphics[width=0.19\linewidth]{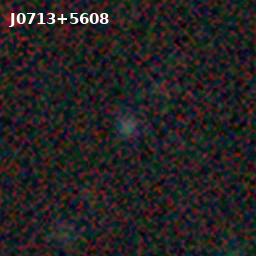}
    \includegraphics[width=0.19\linewidth]{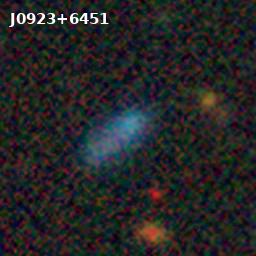}
    \includegraphics[width=0.19\linewidth]{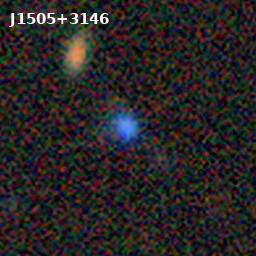}
    \includegraphics[width=0.19\linewidth]{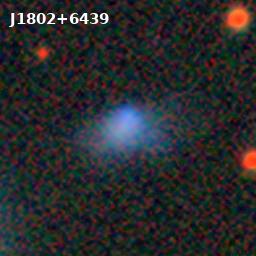}
    \includegraphics[width=0.19\linewidth]{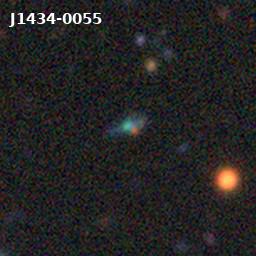}
\end{minipage}
\begin{minipage}{1.00\textwidth}
    \includegraphics[width=0.19\linewidth]{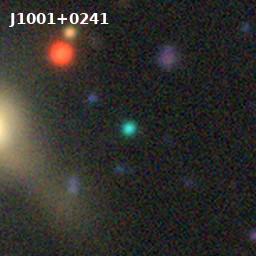}
    \includegraphics[width=0.19\linewidth]{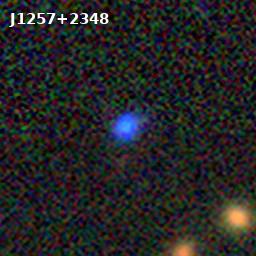}
    \includegraphics[width=0.19\linewidth]{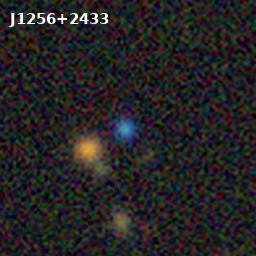}
    \includegraphics[width=0.19\linewidth]{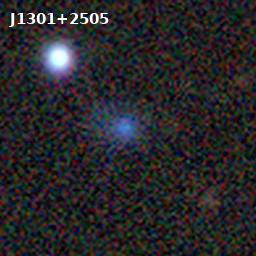}
    \includegraphics[width=0.19\linewidth]{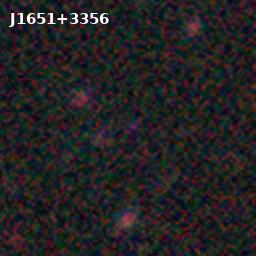}
\end{minipage}
\begin{minipage}{1.00\textwidth}
    \includegraphics[width=0.19\linewidth]{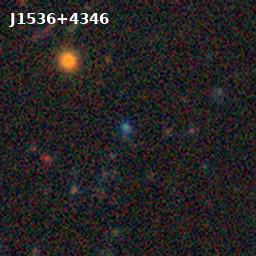}
\end{minipage}
\caption[short]{Color composite image of galaxies made using three optical bands (g, r, z) taken from the Legacy Survey DR9 sky map, except for J0224-0328, J1212-0044, J1003+0123, J1434-0055, and J1001+0241, which incorporate an image from Hyper Suprime-Cam (HSC) Subaru Strategic Program DR3. Each image is sized uniformly at 25" by 25" centered on the spectral target of all observed objects. North is up and east is left. The image of J1225+3151 has artifacts from a bright star nearby.
}
\label{figure:images}
\end{figure*}

\begin{figure}
\resizebox{1.00\hsize}{!}{\includegraphics[angle=000]{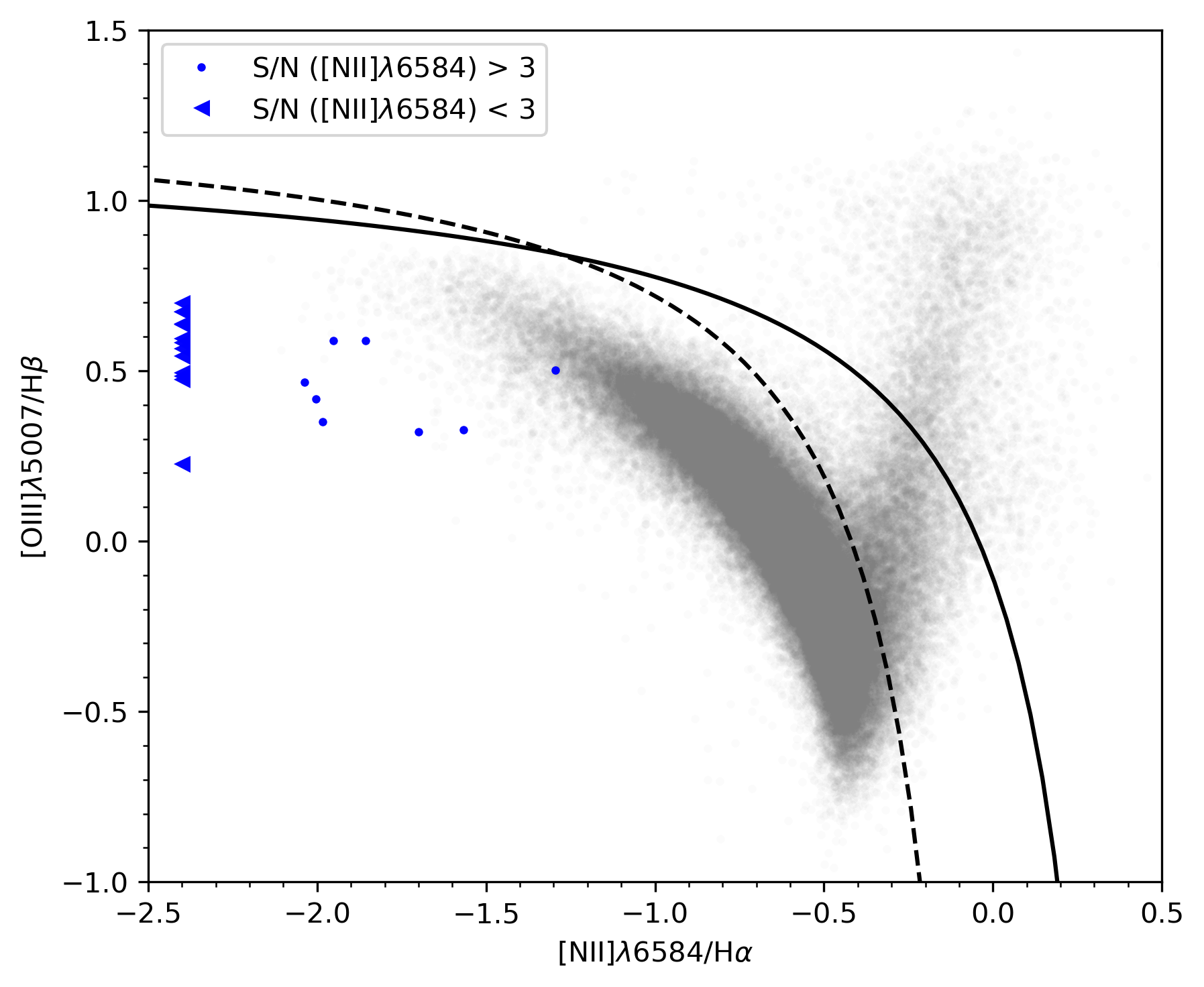}}
\caption{%
BPT diagram for our sample of extremely low-metallicity galaxies. Blue circles show galaxies with detected \ion{N}{II}]$\lambda$6584 line, while blue triangles represent galaxies with S/N < 3 in \ion{N}{II}]$\lambda$6584. Gray circles represent the entire sample of DESI galaxies with $z < 0.5$. The solid line is the theoretical upper limit for star-forming galaxies presented by \citet{Kewley2001}. The dashed line is the demarcation line proposed by \citet{Kauffmann2003} as a lower limit for the area of AGNs.
}
\label{figure:BPT}
\end{figure}

\section{Data}
\label{sect:data}

For this study, we analyzed a subset of galaxies from the DESI EDR survey \citep{DESI2023}. The DESI EDR includes spectra for 1.8 million distinct targets observed during Survey Validation, commissioning, and special tiles, spanning December 2020 to June 2021.

We analyzed 666773 DESI spectra of galaxies with redshifts of less than 0.5, where H$\alpha$ falls within the DESI wavelength range. From the analysis point of view, DESI spectra do not show significant differences with respect to the other optical spectroscopic surveys (e.g., the Sloan Digital Sky Survey (SDSS), the Calar Alto Legacy Integral Field Area Survey (CALIFA), Mapping Nearby Galaxies at APO (MaNGA)), as they have compatible spatial and spectral resolutions and wavelength ranges. Therefore, for analysis of DESI spectra, we employed the methodology that has been applied in our previous studies of the multi-object spectroscopy (MOS) and integral field units (IFU) spectra outlined in, for example, \citet{Zinchenko2016,Zinchenko2021} and \citet{LaraLopez2023}. In essence, the stellar background was fitted using the public version of the STARLIGHT code~\citep{CidFernandes2005,Mateus2006,Asari2007}, customized for parallel processing of individual spectra. Simple stellar population (SSP) spectra from the evolutionary synthesis models by \citet{BC03} were used to fit the stellar spectra and were subsequently subtracted from the observed spectrum to derive a pure gas spectrum.

We used our ELF3D code for emission line fitting, which is able to fit complex line profiles such as blends or profiles with broad and narrow components. The recent ELF3D version is based on the LMfit library and shared-memory parallelization. This approach reduces the number of dependencies and improves the portability of the code. In this work, we fit each emission line with a single Gaussian profile. A full list of measured emission lines can be found in Table~\ref{table:fluxes}. A comparison of measured fluxes with the ones obtained from SDSS spectra in the MPA/JHU catalog is discussed in Appendix~\ref{sect:comparison-flux} and Fig.~\ref{figure:flux-compare}.

Then, the line fluxes were corrected for interstellar reddening following the extinction curve from \citet{Fitzpatrick1999}, assuming a Balmer line ratio for high electron temperatures (T$_e \sim 20000$~K), $\text{H}\alpha/\text{H}\beta = 2.75$. In cases where the measured value of $\text{H}\alpha/\text{H}\beta$ was less than 2.75, the reddening correction was set to zero.

To select spectra of star-forming regions, we applied the $\log$([\ion{O}{III}]$\lambda$5007/H$\beta$) -- $\log$([\ion{N}{II}]$\lambda$6584/H$\alpha$) diagram~\citep{BPT}. Consequently, we selected spectra where [\ion{O}{III}]$\lambda$5007, H$\beta$, and H$\alpha$ emission lines were detected with a signal-to-noise ratio (S/N) exceeding 3. No S/N constraint was imposed on [\ion{N}{II}]$\lambda$6584, because this line is anticipated to exhibit very low flux in \ion{H}{II}~regions with extremely low metallicity~\citep[e.g.,][]{Denicolo2002}. When the S/N of [\ion{N}{II}]$\lambda$6584 fell below 3, the spectrum was classified as belonging to a star-forming region. In instances of detected [\ion{N}{II}]$\lambda$6584, we selected only spectra where the primary ionization source was identified as massive stars, guided by the dividing line proposed by \cite{Kauffmann2003}. Additionally, we imposed $\text{S/N} > 3$ for [\ion{O}{II}]$\lambda\lambda$3727,3729 lines to enable estimation of the chemical abundance of the O$^+$ ion. 

The relation between the auroral line [\ion{O}{III}]$\lambda$4363 and the electron temperature is significantly nonlinear and becomes flatter at high electron temperatures, which corresponds to the low metallicity. Therefore, small uncertainties in [\ion{O}{III}]$\lambda$4363 flux lead to high uncertainties in the electron temperature, and objects with slightly overestimated [\ion{O}{III}]$\lambda$4363 flux will have underestimated metallicity. To minimize this effect and the probability of the spurious detection of the [\ion{O}{III}]$\lambda$4363 line, we adopted a slightly higher detection threshold of $\text{S/N} > 4$ for the auroral line [\ion{O}{III}]$\lambda$4363. It is interesting to note that applying this criterion to the [\ion{O}{III}]$\lambda$4363 line leads to a final sample of galaxies (described in Sect.~\ref{sect:analysis}) with $\text{S/N} > 5$ for all [\ion{O}{II}]$\lambda\lambda$3727,3729, H$\beta$, [\ion{O}{III}]$\lambda$5007, and H$\alpha$ lines.

\section{Determination of the chemical abundances, electron densities, and temperatures}
\label{sect:abundance}

In this section, we describe our methodology for determining electron densities (n$_e$) and temperatures (T$_e$) along with chemical abundances. Computations were executed using PyNeb code version 1.1.18~\citep{Luridiana2015} with default atomic data PYNEB\_23\_01.

The spectral resolution of DESI spectra allows robust measurements of each line in the [\ion{O}{II}]$\lambda\lambda$3727,3729 doublet. Consequently, we derived n$_e$ from the [\ion{O}{II}]$\lambda$3727/[\ion{O}{II}]$\lambda$3729 and [\ion{S}{II}]$\lambda$6717/[\ion{S}{II}]$\lambda$6731 diagnostic line ratios. The diagnostic line ratios of both \ion{O}{II} and \ion{S}{II} doublets consistently indicate a low-density regime (n$_e < 300$~cm$^{-3}$) for all objects in our sample. Due to substantial uncertainty at low densities, the computed n$_e$ values were not used to determine T$_e$ or chemical abundances. Instead, a fixed value of n$_e = 100$~cm$^{-3}$ was assumed, because its effect is negligible in the calculation of final ionic abundances.

For all galaxies within our sample, T$_e$ values were directly derived using the [\ion{O}{III}]$\lambda$4363 auroral line. Following the approach described by \citet{PerezMontero2017} and \citet{Garnett1992}, we adopted a three-zone temperature model with electron temperatures $t_h$, $t_m$, and $t_l$ in the inner, intermediate, and outer zones, respectively, which establishes a link between T$_e$ derived in the [\ion{O}{III}] zone and the high-temperature zone:
\begin{equation}
    t_h = T_e([\ion{O}{III}])
,\end{equation}
and assumes the same $t_h$ in zones of [\ion{Ne}{III}], [\ion{Fe}{III}], and [\ion{Ar}{IV}]. For other ions, we adopted the following scheme:
\begin{equation}
    t_m = T_e([\ion{Ar}{III}]) = T_e([\ion{S}{III}])
,\end{equation}
\begin{equation}
    t_l = T_e([\ion{O}{II}]) = T_e([\ion{N}{II}]) = T_e([\ion{S}{II}])
.\end{equation}
We then derived t$_m$ from the relation between T$_e$([\ion{S}{III}]) and T$_e$([\ion{O}{III}]) proposed by \citet{Hagele2006}: 
\begin{equation}
t_m   =  1.19 t_h - 0.32,
\label{equation:tmth}
\end{equation}
and t$_l$ was obtained from the relation between $t_l$ and $t_h$ suggested by \citet{Garnett1992}:
\begin{equation}
t_l   =  0.7 t_h + 0.3.
\label{equation:tlth}
\end{equation}

Having determined n$_e$ and T$_e$, we proceeded to derive the abundance of the O$^+$ ion from the flux of [\ion{O}{II}]$\lambda\lambda$3727,3729 lines, the abundance of the O$^{2+}$ ion from [\ion{O}{III}]$\lambda\lambda$4959,5007 lines, and the total oxygen abundance as the sum of the ionic abundances of O$^{+}$/H$^{+}$ and O$^{2+}$/H$^{+}$. This procedure was applied to other ions for which we detected at least one nebular emission line. Then, ionization correction factors (ICFs) from \citet{Izotov2006} were applied to account for the unseen ionization states.

\section{Chemical abundances of galaxies with extremely low metallicity}
\label{sect:analysis}

We derived the oxygen abundance using the T$_e$ method for a sample of galaxies obtained in Sect.~\ref{sect:data}. From this sample, we selected a final sample of galaxies with 12+$\log(\rm{O/H}) < 7.3$, comprising 21 objects. Among them, 7 galaxies were recently reported as extremely metal-poor by \citet{Zou2024}, while the metallicity of the other 14 galaxies is measured for the first time.
The general characteristics of these 21 selected galaxies are detailed in Table~\ref{table:info}, where information obtained from the DESI catalog is presented, including the name, DESI target ID, right ascension (RA) and declination (DEC) coordinates, redshift, magnitudes in $g$ and $r$ bands, and exposure time of the DESI coadd spectrum. Additionally, we estimated the absolute magnitude $M_g$ based on redshift and $g$ magnitude. Color composite image cutouts for each galaxy are provided in Fig.~\ref{figure:images}.

When corresponding emission lines are detected in the spectrum, we derived values of electron density, electron temperature, and chemical abundances of O, N, Ne, Ar, and S, which are summarized in Table~\ref{table:abundance}. Furthermore, Table~\ref{table:ions} encapsulates the chemical abundances of the individual ions listed above.
Emission line intensities and extinction coefficients C(H$\beta$) used for calculations and equivalent widths of H$\beta$ EW(H$\beta$) are reported in the Appendix in Table~\ref{table:fluxes}.

The BPT diagram in Fig. \ref{figure:BPT} illustrates that our sample of DESI galaxies predominantly occupy the region characterized by high [\ion{O}{III}]$\lambda$5007/H$\beta$ and low [\ion{N}{II}]$\lambda$6584/H$\alpha$, indicative of low-metallicity \ion{H}{II}~regions. In the luminosity--metallicity diagram in Fig.\ref{figure:LZ}, most galaxies in our sample exhibit metallicities lower than the average values for low-$z$ SDSS galaxies with comparable luminosity \citep{Guseva2009}, aligning more closely with the luminosity--metallicity relation for local metal-poor extreme blue compact galaxies (XBCGs) identified by \citet{KunthOstlin2000}. According to \citet{KunthOstlin2000}, XBCGs are 0.5~dex less metal rich at a given luminosity compared to the luminosity--metallicity relation for low-mass galaxies. The large number of galaxies in our sample with properties similar to XBCGs may be explained by the DESI sample selection. The DESI survey was designed to map the large-scale structure over a wide range of look-back times. Therefore, the sample selection favored compact objects, presumably at higher redshifts, rather than \ion{H}{II} regions in nearby galaxies.

\subsection{N/O abundance ratio}

The nitrogen line [\ion{N}{II}]$\lambda$6584 has been detected in eight galaxies from our sample, which allowed us to derive the N/O ratio.
The top panel of Fig.~\ref{figure:oh-no} presents the O/H--N/O diagram for our sample in comparison with a reference sample of \ion{H}{II}~regions in nearby galaxies and BCDs/BCGs gathered from the literature~\citep{Annibali2019,Croxall2016,Haurberg2015,Nicholls2014,Toribio2017,Esteban2009,Fricke2001,Garnett1997,GonzalezDelgado1994,Guseva2000,Guseva2001,Guseva2003a,Guseva2003b,Guseva2004,Guseva2011,Hagele2008,Hirschauer2015,Izotov1994,Izotov1997,Izotov1998,Izotov1998b,Izotov1999,Izotov2001,Izotov2004,Izotov2004b,Izotov2009,Izotov2012,Kehrig2004,Kennicutt2001,Kniazev2000,Lee2005,Lee2003,Lee2004,Magrini2009,Melbourne2004,Melnick1992,Pagel1992,Pena2007,Pustilnik2002,Pustilnik2003a,Pustilnik2003b,Pustilnik2005,Pustilnik2006,Saviane2008,Skillman1993,Skillman2003,Stanghellini2010,Terlevich1991,Thuan1995,Thuan1999,vanZee1997,vanZee2000,Vilchez2003}. This reference sample has been created to collect spectra with reliable measurements of auroral lines published over the last decade. For both samples, we applied the method for determining chemical abundances described in Sect.~\ref{sect:abundance}. The red dashed line on the plot denotes the median N/O ratio for the reference sample with an oxygen abundance of less than 8.0~dex. Remarkably, five out of the eight DESI galaxies with extremely low oxygen abundance exhibit a higher N/O ratio than the median value for the reference sample. Their 1$\sigma$ confidence intervals of the N/O ratio lie above the median N/O ratio for the reference sample, $\log(\rm{N/O}) = -1.48$, while for the remaining three galaxies, the 1$\sigma$ confidence intervals of the N/O ratio overlap with the median N/O ratio for the reference sample. It should be noted that there is no significant difference between the N/O ratio of BCDs/BCGs and the rest of the reference sample, although we note a mild elevation of the N/O ratio for BCDs/BCGs in a narrow range of the oxygen abundance around 7.2~dex. However, BCDs/BCGs with higher and lower metallicity do not show such a N/O elevation.

Enhancement of the N/O ratio at low metallicity has been reported in several works \citep[e.g.,][]{Guseva2011,PerezMontero2011,Kumari2018,Zinchenko2022}.
As has been discussed in \citet{PerezMontero2011} and \citet{Kumari2018}, several mechanisms can be responsible for an enhanced N/O ratio at low metallicity, including a localized nitrogen enrichment of the interstellar medium by WR stars, an inflow of low-metallicity gas, and varying star formation efficiency (SFE). We did not find WR bumps in the spectra of galaxies from our sample. However, since galaxies in our sample are very faint (majority below 22 mag), the S/N ratio in the continuum is quite low for most galaxies. Thus, the presence of WR stars cannot be excluded and additional observations with higher S/N ratios in the background near possible WR bumps are needed to confirm or rule out WR stars as a source of enhanced N/O ratio in these galaxies, though massive WR stars are rare at very low metallicity.

An inflow of low-metallicity gas is very unlikely to result in a measured N/O enhancement in extremely metal-poor galaxies, because it assumes very significant dilution of their O/H from 12+$\log(\rm{O/H}) \sim 8.2$, where N/O starts to increase with O/H, down to 12+$\log(\rm{O/H}) \sim 7.0-7.3$. For example, dilution of 12+$\log(\rm{O/H}) = 8.2$ gas by 7.0~dex gas to obtain 7.2~dex from 8.2~dex would require about
25 times more inflow gas than the in situ gas with an initial metallicity of 8.2~dex.

\begin{figure}
\resizebox{1.00\hsize}{!}{\includegraphics[angle=000]{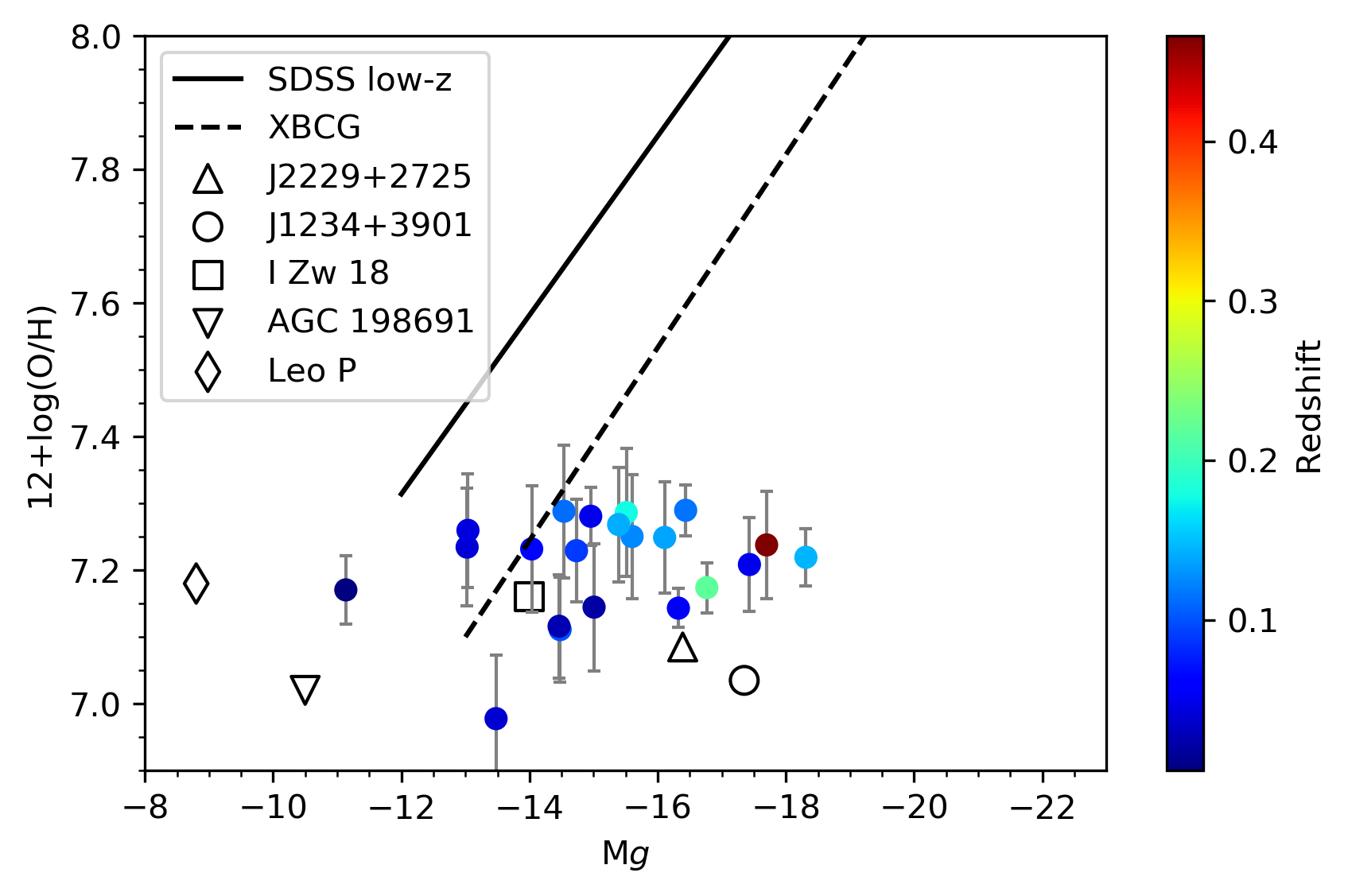}}
\caption{%
Luminosity--metallicity diagram for our sample of extremely low-metallicity galaxies. Colored circles show galaxies from our sample with color-coded redshift. The solid line is the best fit of the luminosity--metallicity relation for SDSS low-z galaxies obtained by \citet{Guseva2009}. The solid line is the best fit of the luminosity--metallicity relation for XBCG obtained by \citet{KunthOstlin2000}. Open symbols represent galaxies with extremely low metallicity collected from the literature.
}
\label{figure:LZ}
\end{figure}

In this regard, the opposite scenario is more likely. The inflow of even a small amount of high-metallicity gas, which also has a high N/O ratio, may significantly elevate the N/O ratio. For example, mixing pristine gas of 12+$\log(\rm{O/H}) = 7.0$ and $\log(\rm{N/O}) = -1.5$ with only 1\% (by atom number) of gas with solar metallicity increases the oxygen abundance by $\sim 0.2$~dex and N/O by $\sim 0.3$~dex. Although this scenario has not been widely discussed for extremely low-metallicity galaxies, it has been suggested for more metal-reach galaxies \citep[see, e.g.,][]{Schaefer2020,Zinchenko2023}. Such contamination by metal-rich gas may also explain the scatter in the N/O ratio, as different galaxies have different gas-accretion histories. 
Discussing the origin of XBCGs, \citet{KunthOstlin2000} specify that XBCGs are likely the product of mergers. As can be seen from Fig.~\ref{figure:LZ}, most of the galaxies from our sample better fit the luminosity--metallicity relation for XBCGs from \citet{KunthOstlin2000} than they do the luminosity--metallicity
relation of the SDSS galaxies from \citet{Guseva2009}. This fact supports the scenario of inflow of gas with higher metallicity either as an independent event or as a result of merging.

\subsection{Ne/O, Ar/O, and S/O abundance ratios}

Enhancement of the abundances of other chemical elements in metal-poor star-forming regions also remains an open question. Recent theoretical and observational studies have reported enhanced abundances of some chemical elements in extremely metal-poor star-forming dwarf galaxies, where 12+log(O/H) $\le$ 7.69 \citep[e.g.,][]{Takahashi2018,Kojima2021,Goswami2022,Goswami2024}. These studies attribute the overabundance of certain chemical elements at very low metallicity to the explosion of very massive stars as pair-instability supernovae (PISNe).

In our sample of 21 DESI galaxies, we derived Ne/O in 20 galaxies, Ar/O in 9 galaxies, and S/O ratios in 8 galaxies. We present these measurements in Fig.~\ref{figure:oh-no} following the same approach as for N/O. Similar to previous findings by \citet{Kojima2021} and \citet{ArellanoCordova2024}, we find no enhancement of Ne/O in comparison to the values calculated for low-metallicity galaxies from the literature. However, the majority of galaxies in our sample exhibited increased Ar/O ratios, in contrast to the results reported by \citet{Kojima2021} and \citet{ArellanoCordova2024}. Additionally, we observed an increase in the S/O ratio in most galaxies from our sample.

It is noteworthy that the behavior of the S/O ratio with respect to metallicity has been a longstanding open question. Some studies have indicated that the S/O ratio remains constant with metallicity \citep[e.g.,][]{Izotov2006,Berg2020,ArellanoCordova2024}, while others have suggested an increase in the S/O ratio at low metallicity \citep[e.g.,][]{Vilchez1988,Dors2016,Diaz2022}. Recently, \citet{Goswami2024} showed that very massive stars undergoing the PISN stage may play a significant role in the chemical enrichment of sulfur in metal-poor galaxies.

\section{J0713+5608: galaxy with the lowest oxygen abundance in our sample}
\label{sect:J0713+5608}

\begin{figure}
\resizebox{1.00\hsize}{!}{\includegraphics[angle=000]{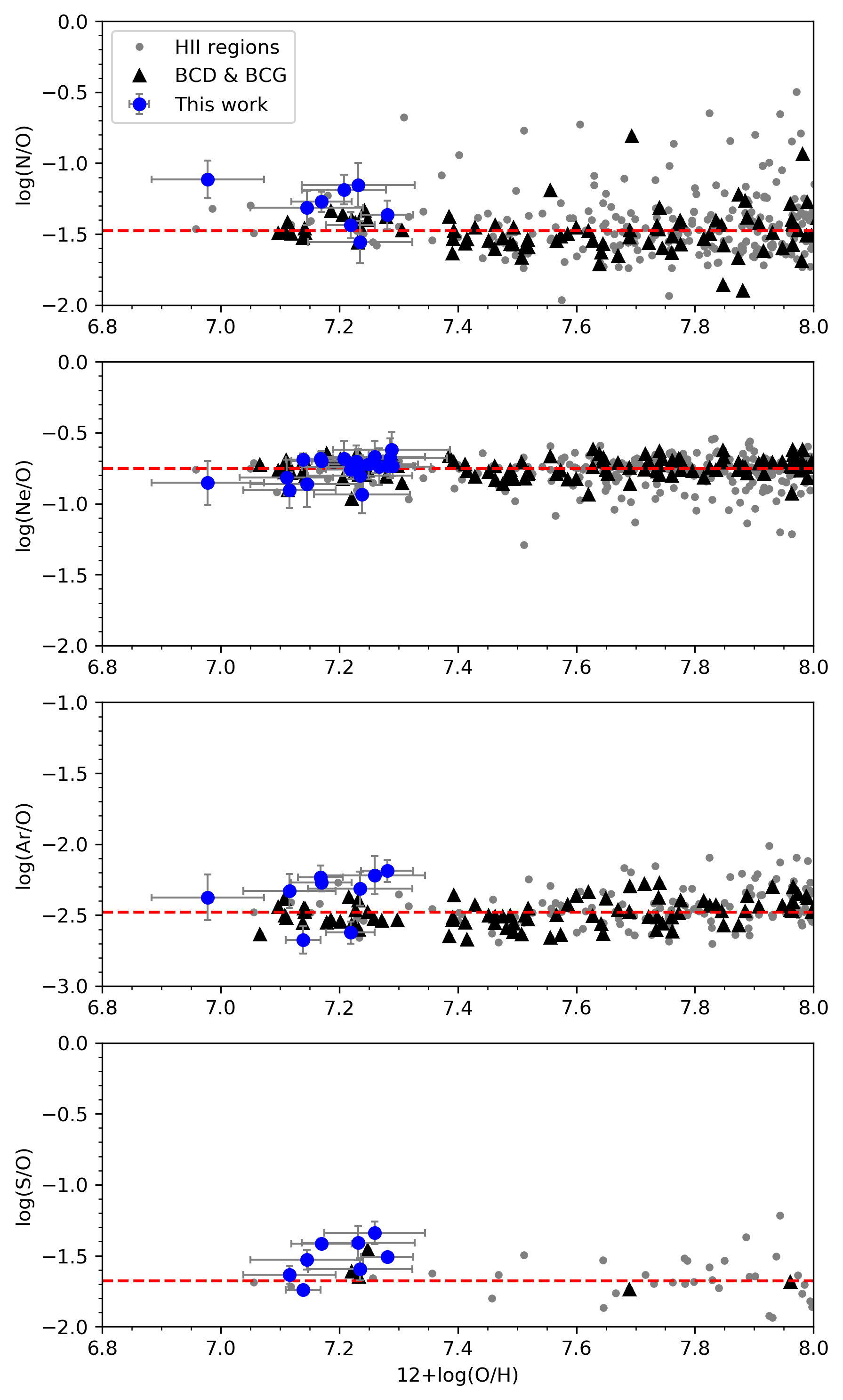}}
\caption{%
N/O, Ne/O, Ar/O, and S/O ratios with respect to oxygen abundance. Blue circles are our sample of extremely metal-poor galaxies. Gray points and black triangles represent a reference sample of \ion{H}{II}~regions and BCD/BCG galaxies, respectively, with measured oxygen auroral line collected from the literature. Red dashed lines are medians of N/O, Ne/O, Ar/O, and S/O for the reference \ion{H}{II}~regions with O/H in the range 7.0 -- 8.0~dex.
}
\label{figure:oh-no}
\end{figure}

One of the galaxies in our sample, J0713+5608, exhibits an exceptionally low oxygen abundance of 6.978$\pm$0.095dex, lower than the oxygen abundance of 7.11 -- 7.17~dex observed in the well-known extremely metal-poor galaxy I~Zw~18~\citep{Izotov1998,Kehrig2016}. This measurement is compatible with the previously reported lowest oxygen abundance in galaxies, such as DDO~68~\citep{Annibali2019} and J0811+4730~\citep{Izotov2018}. Despite its minimal oxygen abundance, J0713+5608 demonstrates the highest N/O ratio among our DESI galaxy sample, as depicted in Fig.\ref{figure:oh-no}. In contrast, its Ne/O and Ar/O ratios align with the typical values observed in our sample and are similar to the median values observed in the reference \ion{H}{II}~regions.

In the DESI Legacy Survey image, J0713+5608 appears as a faint BCG and occupies a corresponding position on the luminosity--metallicity diagram. Remarkably, with an absolute magnitude of M$_g = -13.48$, J0713+5608 lies below the average luminosity--metallicity relation identified by \citet{KunthOstlin2000} for XBCGs.

\section{Conclusions}
\label{section:Summary}

We analyzed a subset of galaxies from the DESI EDR survey \citep{DESI2023} to identify extremely metal-poor galaxies. By focusing on DESI spectra with redshifts below 0.5, allowing the inclusion of H$\alpha$ within the DESI wavelength range, we performed measurements of the auroral line [\ion{O}{III}]$\lambda$4363 and calculated the oxygen abundance using the direct T$_e$ method.

Among 666773 analyzed galaxies, we find that 21 galaxies exhibit oxygen abundances of below 7.3~dex. Alongside our determination of the oxygen abundance, we derived N/O, Ne/O, Ar/O, and S/O ratios for some of these galaxies, which we then compared with corresponding abundance ratios obtained for a reference sample of nearby \ion{H}{II}~regions and BCD/BCG galaxies. These reference values were derived from a dataset of measured oxygen auroral lines collected from the literature. We find that a fraction of DESI galaxies with extremely low oxygen abundance exhibit higher N/O, Ar/O, and S/O ratios in comparison to the median values observed within the reference sample. However, it is worth noting that the Ne/O ratio of these extremely metal-poor DESI galaxies does not exhibit a substantial deviation from the Ne/O ratio observed in the reference low-metallicity \ion{H}{II}~regions and BCD/BCG galaxies. We also suggest that the elevated N/O ratio in some galaxies from our sample can be explained by the scenario of the inflow of gas with higher metallicity either as an independent event or as a result of a merger. However, contributions from WR stars also cannot be excluded.

The galaxy J0713+5608 stands out in our study because of its remarkably low oxygen abundance of 6.978$\pm$0.095~dex. This measurement aligns with the lowest known oxygen abundances in galaxies to date. Given the relatively high uncertainty, this galaxy may have the lowest oxygen abundance ever found. Additionally, we found J0713+5608 to exhibit an enhanced N/O ratio compared to the typical N/O ratio observed in metal-poor galaxies within the local Universe.

\begin{acknowledgements}

We are grateful to the referee, A. Hirschauer, for his constructive comments. 
We thank Evan Skillman for useful comments.\\
JVM and CK acknowledge financial support from the Spanish MINECO grant PID2022-136598NB-C32 and from
the AEI  “Center of Excellence Severo Ochoa” award to the IAA (SEV-2017-0709).\\
This research used data obtained with the Dark Energy Spectroscopic Instrument (DESI). DESI construction and operations are managed by the Lawrence Berkeley National Laboratory. This material is based upon work supported by the U.S. Department of Energy, Office of Science, Office of High-Energy Physics, under Contract No. DE–AC02–05CH11231, and by the National Energy Research Scientific Computing Center, a DOE Office of Science User Facility under the same contract. Additional support for DESI was provided by the U.S. National Science Foundation (NSF), Division of Astronomical Sciences under Contract No. AST-0950945 to the NSF’s National Optical-Infrared Astronomy Research Laboratory; the Science and Technology Facilities Council of the United Kingdom; the Gordon and Betty Moore Foundation; the Heising-Simons Foundation; the French Alternative Energies and Atomic Energy Commission (CEA); the National Council of Science and Technology of Mexico (CONACYT); the Ministry of Science and Innovation of Spain (MICINN), and by the DESI Member Institutions: www.desi.lbl.gov/collaborating-institutions.

\end{acknowledgements}

\bibliographystyle{aa}
\bibliography{reference}

\begin{appendix}

\section{Comparison with the MPA/JHU catalog}
\label{sect:comparison-flux}

\begin{figure*}
\resizebox{0.90\hsize}{!}{\includegraphics[angle=000]{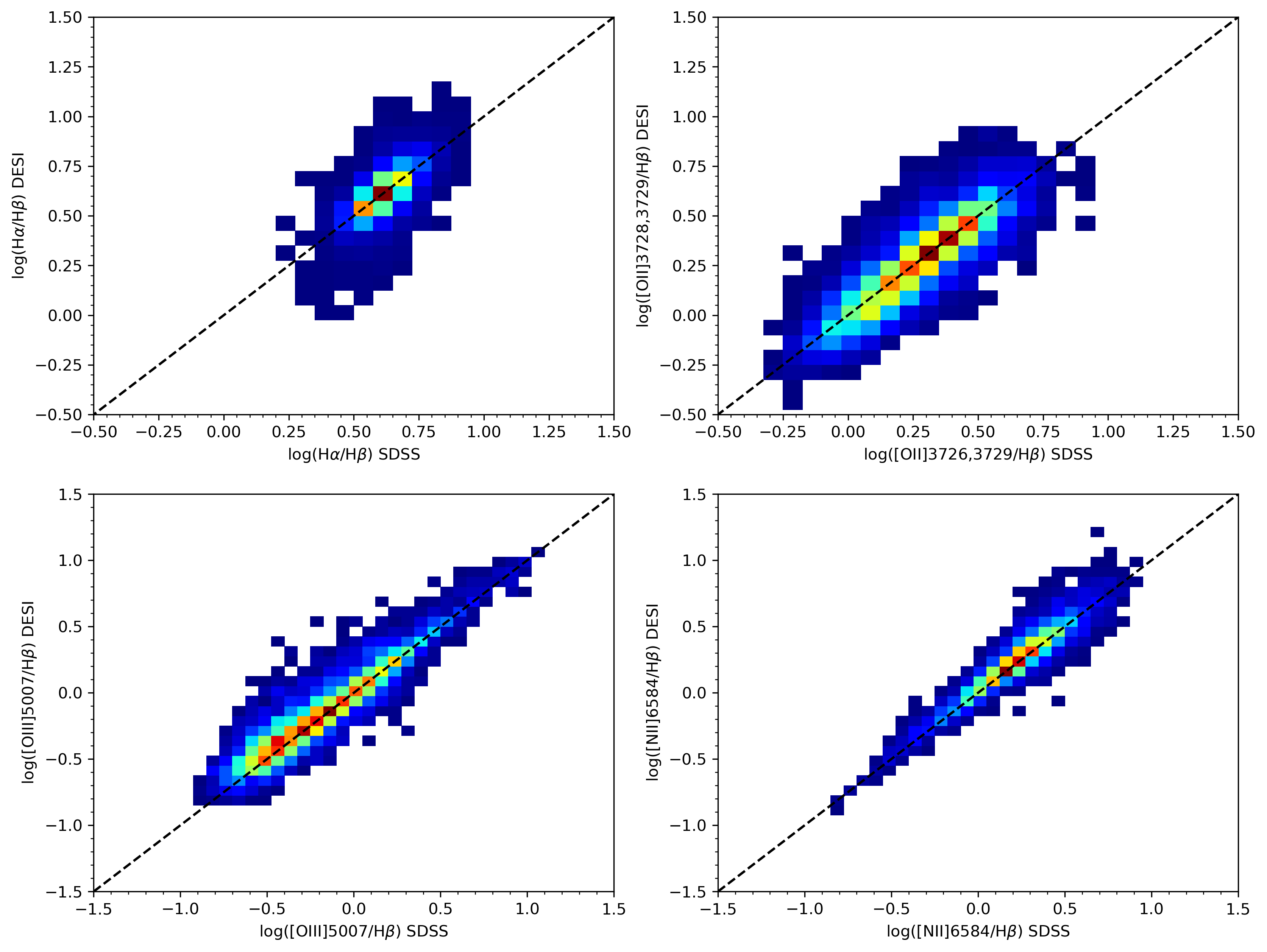}}
\caption{%
Comparison of the emission line fluxes measured by us for the DESI galaxies with the ones presented in the MPA/JHU catalog as a part of SDSS. 2D histograms show the number of objects in the flux bin. Only bins with 3 or more objects are shown. The dashed line shows one-to-one correspondence between fluxes.
}
\label{figure:flux-compare}
\end{figure*}

To check the reliability of our measurements of the flux of emission lines in DESI galaxies, we performed a comparison of the emission line fluxes used in this work with the ones presented in the MPA/JHU catalog~\footnote{https://www.sdss4.org/dr17/spectro/galaxy\_mpajhu/} for the SDSS \citep{Brinchmann2004}. From both catalogs we selected spectra with $\text{S/N} > 3$ in [\ion{O}{II}]$\lambda\lambda$3727,3729, H$\beta$, \ion{N}{II}]$\lambda$6584, [\ion{O}{III}]$\lambda$5007, and H$\alpha$ lines. Then, we cross-matched their positions on the sky requiring the maximal difference in fiber positions of 0.3~arcsecond which is significantly smaller than the diameter of DESI and SDSS fibers.
After applying these criteria, we found 5374 objects with emission line fluxes available in our DESI catalog and MPA/JHU catalog for SDSS galaxies. 

In Fig.~\ref{figure:flux-compare} we compared the fluxes normalized by H$\beta$ obtained from our catalog of DESI objects and the MPA/JHU catalog. It demonstrates that our measurements of the nebular emission lines [\ion{O}{II}]$\lambda\lambda$3727,3729, [\ion{O}{III}]$\lambda$5007, [\ion{N}{II}]$\lambda$6584, needed for determination of the oxygen abundance and N/O ratio, are in a very good agreement with the ones from the MPA/JHU catalog. H$\alpha$/H$\beta$ line ratio used for extinction correction also shows no systematic offset between our measurements and the MPA/JHU catalog.

\section{Emission line intensities}

\begin{landscape}
\begin{table}
\centering
\caption{Emission line intensities before correction for extinction. Fluxes are in units of 10$^{-17}$~erg~s$^{-1}$~cm$^{-2}$}.
\label{table:fluxes}
\begin{tabular}{lccccccccccccccccccccc}
\hline\hline
{}        & J1143-0139 & J0941+3209 & J1333+3326 & J0224-0328 & J1212-0044 & J1003+0123 & J1225+3151 & J0847+3257 & J0901+3336 & J0651+3843 & J0713+5608 \\
\hline
{[\ion{O}{II}]$\lambda$3726}  & 58.11$\pm$2.97&52.52$\pm$4.89&   8.13$\pm$2.68& 6.05$\pm$0.92& 20.74$\pm$1.93& 9.63$\pm$0.75&  3.81$\pm$1.23&  6.20$\pm$1.64& 18.95$\pm$2.08&  4.22$\pm$1.17& 10.85$\pm$1.20&  \\
{[\ion{O}{II}]$\lambda$3729}  &76.68$\pm$3.16 &79.85$\pm$5.07&  14.33$\pm$2.91& 9.12$\pm$0.97& 28.57$\pm$1.98&11.83$\pm$0.77&  9.76$\pm$1.49&  8.06$\pm$1.73& 29.92$\pm$2.22&  8.23$\pm$1.33& 15.10$\pm$1.30&  \\
H9                            &          ---  &         --- &             --- &          --- &           --- &          --- &  5.14$\pm$1.02&           --- &           --- &           --- &           --- &  \\
{[\ion{Ne}{III}]$\lambda$3869}& 44.47$\pm$3.60&13.99$\pm$3.91&            --- & 5.76$\pm$0.91& 10.00$\pm$1.53& 3.85$\pm$0.62& 10.38$\pm$1.15& 10.45$\pm$1.79& 18.67$\pm$2.04&  6.62$\pm$1.03&  4.74$\pm$0.84&  \\
{[\ion{He}{I}]}+H8            &29.38$\pm$3.01 &18.07$\pm$3.72&            --- & 3.01$\pm$0.79&  4.56$\pm$1.39& 2.18$\pm$0.57&  5.97$\pm$1.09&           --- &  7.50$\pm$1.87&  3.87$\pm$0.89&  5.61$\pm$0.86&  \\
{[\ion{Ne}{III}]$\lambda$3967}& 14.17$\pm$2.50&          --- &            --- &          --- &           --- &          --- &           --- &           --- &  5.38$\pm$1.51&           --- &  2.14$\pm$0.69&  \\
H$\epsilon$                   &28.53$\pm$2.75 &          --- &            --- & 4.83$\pm$1.09&  5.38$\pm$1.17& 7.43$\pm$1.83&  4.12$\pm$0.89&           --- &  8.82$\pm$1.56&           --- &  4.93$\pm$0.79&  \\
{[\ion{He}{I}]$\lambda$4027}  &          ---  &          --- &            --- &          --- &           --- &          --- &           --- &           --- &           --- &           --- &           --- &  \\
{[\ion{S}{II}]$\lambda$4069}  &          ---  &          --- &            --- &          --- &           --- &          --- &           --- &           --- &           --- &           --- &           --- &  \\
{[\ion{S}{II}]$\lambda$4076}  &         ---  &           --- &            --- &          --- &           --- &          --- &           --- &           --- &           --- &           --- &           --- &  \\
H$\delta$                     &40.72$\pm$2.68 &18.87$\pm$2.45&            --- & 4.36$\pm$0.65& 13.11$\pm$1.31& 3.28$\pm$0.73&  6.74$\pm$0.86&  7.44$\pm$1.42& 14.44$\pm$1.72&  7.03$\pm$1.35&  6.09$\pm$0.77&  \\
H$\gamma$                     &80.34$\pm$2.70 &31.93$\pm$2.02&   9.41$\pm$1.45& 7.51$\pm$0.53& 20.87$\pm$1.17& 7.26$\pm$0.53& 13.64$\pm$0.95& 13.53$\pm$1.08& 23.47$\pm$1.02&  9.43$\pm$0.89& 12.94$\pm$0.75&  \\
{[\ion{O}{III}]$\lambda$4363} &19.54$\pm$1.87 & 6.40$\pm$1.49&   6.36$\pm$1.31& 1.90$\pm$0.44&  4.27$\pm$0.95& 2.22$\pm$0.46&  4.01$\pm$0.77&  4.79$\pm$0.91&  7.89$\pm$0.69&  3.21$\pm$0.75&  2.80$\pm$0.61&  \\
{[\ion{He}{I}]$\lambda$4472}  &          ---  &         --- &             --- &          --- &           --- &          --- &           --- &           --- &           --- &           --- &           --- &  \\
{[\ion{Fe}{III}]$\lambda$4658}&           --- &          --- &            --- &          --- &           --- &          --- &           --- &           --- &           --- &           --- &           --- &  \\
{[\ion{He}{II}]$\lambda$4686} &          ---  &          --- &            --- &          --- &           --- &          --- &           --- &           --- &           --- &           --- &           --- &  \\
{[\ion{He}{I}]$\lambda$4714}  &          ---  &          --- &            --- &          --- &           --- &          --- &           --- &           --- &           --- &           --- &           --- &  \\
{[\ion{Ar}{IV}]$\lambda$4741} &          ---  &         --- &             --- &          --- &           --- &          --- &           --- &           --- &           --- &           --- &           --- &  \\
H$\beta$                      &176.82$\pm$3.75&61.29$\pm$2.22&  24.47$\pm$1.78&16.88$\pm$0.61& 42.57$\pm$1.01&12.99$\pm$0.53& 25.21$\pm$0.90& 29.50$\pm$1.60& 52.94$\pm$1.25& 18.87$\pm$0.63& 25.96$\pm$0.58&  \\
{[\ion{He}{I}]$\lambda$4923}  &          ---  &         --- &             --- &          --- &           --- &          --- &           --- &           --- &           --- &           --- &           --- &  \\
{[\ion{O}{III}]$\lambda$4959} &184.29$\pm$3.68&48.16$\pm$2.14&  48.87$\pm$2.40&17.14$\pm$0.54& 40.49$\pm$0.98&14.07$\pm$0.54& 35.54$\pm$1.00& 37.47$\pm$1.83& 70.33$\pm$1.76& 26.89$\pm$0.75& 20.26$\pm$0.56&  \\
{[\ion{O}{III}]$\lambda$5007}&518.41$\pm$5.47&130.28$\pm$2.68& 122.22$\pm$2.74&51.45$\pm$0.76&111.45$\pm$1.23&40.52$\pm$0.72& 96.33$\pm$1.37&115.88$\pm$2.91&194.56$\pm$2.48& 73.02$\pm$0.94& 54.35$\pm$0.72&  \\
{[\ion{He}{I}]$\lambda$5017}  &          ---  &          --- &            --- &          --- &           --- &          --- &           --- &           --- &           --- &           --- &           --- &  \\
{[\ion{N}{I}]$\lambda$5199}   &          ---  &         --- &             --- &          --- &           --- &          --- &           --- &           --- &           --- &           --- &           --- &  \\
{[\ion{N}{I}]$\lambda$5201}   &          ---  &          --- &            --- &          --- &           --- &          --- &           --- &           --- &           --- &           --- &           --- &  \\
{[\ion{N}{II}]$\lambda$5755}  &          ---  &          --- &            --- &          --- &           --- &          --- &           --- &           --- &           --- &           --- &           --- &  \\
{[\ion{He}{I}]$\lambda$5875}  &20.50$\pm$1.06 & 6.57$\pm$1.33&            --- &          --- &  3.89$\pm$0.70& 1.38$\pm$0.40&  2.53$\pm$0.48&           --- &  4.54$\pm$0.48&  2.12$\pm$0.47&  3.03$\pm$0.45&  \\
{[\ion{O}{I}]$\lambda$6300}   & 6.08$\pm$0.97 &          --- &            --- &          --- &           --- &          --- &           --- &           --- &  3.13$\pm$0.63&           --- &           --- &  \\
{[\ion{S}{III}]$\lambda$6312} &          ---  &          --- &            --- &          --- &           --- &          --- &           --- &           --- &           --- &           --- &           --- &  \\
{[\ion{N}{II}]$\lambda$6548}  &          ---  &          --- &            --- &          --- &           --- &          --- &           --- &           --- &           --- &           --- &           --- &  \\
H$\alpha$                    &540.63$\pm$4.39&167.81$\pm$2.08&  68.81$\pm$1.73&43.97$\pm$0.79&123.77$\pm$1.23&38.01$\pm$0.50& 66.99$\pm$0.92& 70.85$\pm$1.37&150.07$\pm$1.51& 53.85$\pm$0.55& 70.01$\pm$0.69&  \\
{[\ion{N}{II}]$\lambda$6584}  & 4.96$\pm$1.08 & 4.55$\pm$0.90&            --- &          --- &  1.23$\pm$0.39&          --- &           --- &           --- &           --- &  0.75$\pm$0.25&  1.40$\pm$0.32&  \\
{[\ion{He}{I}]$\lambda$6680}  & 3.49$\pm$0.92 &          --- &            --- &          --- &  1.53$\pm$0.46&          --- &           --- &           --- &  1.29$\pm$0.35&  0.98$\pm$0.28&           --- &  \\
{[\ion{S}{II}]$\lambda$6717}  &11.18$\pm$0.89 &12.09$\pm$0.90&            --- & 1.87$\pm$0.44&  4.54$\pm$0.41& 1.94$\pm$0.30&  1.42$\pm$0.26&  1.89$\pm$0.49&  3.97$\pm$0.57&  1.29$\pm$0.33&  2.40$\pm$0.37&  \\
{[\ion{S}{II}]$\lambda$6731}  & 9.17$\pm$0.81 & 7.13$\pm$0.93&            --- &          --- &  2.79$\pm$0.38&          --- &  0.82$\pm$0.27&           --- &  2.64$\pm$0.51&  1.03$\pm$0.30&  1.81$\pm$0.24&  \\
{[\ion{He}{I}]$\lambda$7067}  & 9.50$\pm$0.79 & 2.38$\pm$0.79&            --- &          --- &  1.22$\pm$0.41&          --- &           --- &           --- &  1.60$\pm$0.49&           --- &           --- &  \\
{[\ion{Ar}{III}]$\lambda$7136}&  3.96$\pm$0.64&          --- &            --- &          --- &  1.97$\pm$0.42&          --- &  1.28$\pm$0.36&           --- &           --- &           --- &  0.73$\pm$0.18&  \\
{[\ion{O}{II}]$\lambda$7320}  & 2.82$\pm$0.76 &          --- &            --- &          --- &           --- &          --- &           --- &           --- &           --- &           --- &           --- &  \\
{[\ion{O}{II}]$\lambda$7330}  & 2.47$\pm$0.74 &          --- &            --- &          --- &           --- & 2.21$\pm$0.55&           --- &           --- &           --- &           --- &           --- &  \\
{[\ion{Ar}{III}]$\lambda$7751}&           --- &          --- &            --- &          --- &           --- &          --- &           --- &           --- &           --- &           --- &           --- &  \\
{[\ion{S}{III}]$\lambda$9068} &          ---  & 4.20$\pm$0.51&            --- &          --- &  3.50$\pm$0.57&          --- &  2.62$\pm$0.40&           --- &           --- &  1.92$\pm$0.61&           --- &  \\
{[\ion{S}{III}]$\lambda$9530} &          ---  &13.01$\pm$1.21&            --- &          --- &           --- &          --- &           --- &           --- &           --- &           --- &           --- &  \\
Pa8                           &          ---  &          --- &            --- &          --- &           --- &          --- &           --- &           --- &           --- &           --- &           --- &  \\
\hline        
C(H$\beta$)               & 0.132$\pm$0.029&0.000$\pm$0.047& 0.028$\pm$0.096&0.000$\pm$0.050&0.069$\pm$0.033&0.078$\pm$0.054&0.000$\pm$0.048&0.000$\pm$0.072&0.038$\pm$0.033&0.046$\pm$0.043&0.000$\pm$0.031&  \\
EW(H$\beta$)                  &  83.4$\pm$1.8 & 51.3$\pm$1.9 &   28.9$\pm$2.1 & 83.1$\pm$3.0 &  63.6$\pm$1.5 & 41.1$\pm$1.7 &  82.3$\pm$3.0 &  94.2$\pm$5.1 &  58.4$\pm$1.4 &  80.1$\pm$2.7 &  35.5$\pm$0.8 &  \\
\hline
\end{tabular}
\end{table}
\end{landscape}

\setcounter{table}{0}

\begin{landscape}
\begin{table}
\centering
\caption{Continued}.
\label{table:fluxes}
\begin{tabular}{lccccccccccccccccccccc}
\hline\hline
{}        &  J0923+6451 & J1505+3146 & J1802+6439 & J1434-0055 & J1001+0241 & J1257+2348 & J1256+2433 & J1301+2505 & J1651+3356 & J1536+4346 \\
\hline
{[\ion{O}{II}]$\lambda$3726}  &   26.24$\pm$2.11& 37.65$\pm$2.70& 140.75$\pm$9.27&  9.88$\pm$1.67&  5.76$\pm$1.51&  14.46$\pm$1.61&  34.83$\pm$3.11&  66.71$\pm$4.99&   5.47$\pm$1.44&   6.30$\pm$1.34 \\
{[\ion{O}{II}]$\lambda$3729}  &   38.56$\pm$2.31& 43.83$\pm$3.08&177.74$\pm$10.77& 12.54$\pm$1.68&  9.43$\pm$1.62&  19.21$\pm$1.83&  54.47$\pm$4.06&  97.03$\pm$6.13&   6.63$\pm$1.46&   6.76$\pm$1.28 \\
H9                            &    8.53$\pm$1.71& 16.59$\pm$1.71&            --- &           --- &  8.09$\pm$1.27&            --- &            --- &            --- &            --- &   4.01$\pm$1.17 \\
{[\ion{Ne}{III}]$\lambda$3869}&   25.03$\pm$1.67& 50.29$\pm$2.69&  62.42$\pm$7.56& 12.83$\pm$1.75& 33.00$\pm$1.65&  13.16$\pm$1.69&  38.46$\pm$3.25&  18.73$\pm$3.64&  13.65$\pm$1.42&   5.35$\pm$1.18 \\
{[\ion{He}{I}]}+H8            &   28.75$\pm$1.74& 30.82$\pm$2.05&  30.83$\pm$6.66&  4.48$\pm$1.45& 16.96$\pm$1.32&   4.64$\pm$1.38&  22.58$\pm$2.66&  26.93$\pm$4.01&   6.23$\pm$1.21&   5.23$\pm$1.22 \\
{[\ion{Ne}{III}]$\lambda$3967}&    6.51$\pm$1.11& 13.85$\pm$1.43&  15.10$\pm$4.94&           --- & 12.65$\pm$1.16&   4.03$\pm$1.21&  11.99$\pm$2.42&  12.96$\pm$2.84&            --- &            ---  \\
H$\epsilon$                   &   20.66$\pm$1.27& 29.14$\pm$1.89&  17.56$\pm$5.02&  5.04$\pm$1.38& 17.72$\pm$1.27&   4.91$\pm$1.28&  16.65$\pm$2.62&  27.81$\pm$3.14&            --- &   6.00$\pm$1.47 \\
{[\ion{He}{I}]$\lambda$4027}  &             --- &           --- &            --- &           --- &           --- &            --- &            --- &            --- &            --- &            ---  \\
{[\ion{S}{II}]$\lambda$4069}  &             --- &           --- &            --- &           --- &           --- &            --- &            --- &            --- &            --- &            ---  \\
{[\ion{S}{II}]$\lambda$4076}  &             --- &           --- &            --- &           --- &           --- &            --- &            --- &            --- &            --- &            ---  \\
H$\delta$                     &   32.94$\pm$1.36& 47.34$\pm$2.41&  37.82$\pm$5.23&  8.78$\pm$1.37& 25.46$\pm$1.35&   7.01$\pm$1.46&  36.25$\pm$2.90&  38.76$\pm$3.21&   9.71$\pm$2.34&   7.94$\pm$1.08 \\
H$\gamma$                     &   57.04$\pm$1.58& 83.74$\pm$1.95&  66.64$\pm$6.02& 15.43$\pm$1.42& 45.56$\pm$1.53&  15.81$\pm$1.40&  55.95$\pm$2.59&  64.74$\pm$3.06&  13.45$\pm$1.45&  10.80$\pm$1.61 \\
{[\ion{O}{III}]$\lambda$4363} &    8.62$\pm$0.93& 21.12$\pm$1.17&  22.69$\pm$4.52&  6.20$\pm$1.13& 14.67$\pm$1.05&   6.51$\pm$1.14&  17.20$\pm$1.54&   8.93$\pm$1.79&   5.30$\pm$1.14&   4.46$\pm$0.79 \\
{[\ion{He}{I}]$\lambda$4472}  &    4.73$\pm$0.93&  5.21$\pm$1.02&            --- &           --- &  5.15$\pm$1.31&            --- &  11.99$\pm$3.76&            --- &   9.13$\pm$2.70&            ---  \\
{[\ion{Fe}{III}]$\lambda$4658}&             --- &           --- &            --- &           --- &           --- &            --- &            --- &            --- &            --- &            ---  \\
{[\ion{He}{II}]$\lambda$4686} &             --- &  3.20$\pm$0.97&            --- &           --- &  3.61$\pm$1.13&            --- &            --- &            --- &            --- &   2.09$\pm$0.66 \\
{[\ion{He}{I}]$\lambda$4714}  &             --- &  3.61$\pm$1.00&            --- &           --- &           --- &            --- &            --- &            --- &            --- &            ---  \\
{[\ion{Ar}{IV}]$\lambda$4741} &             --- &           --- &            --- &           --- &           --- &            --- &            --- &            --- &            --- &            ---  \\
H$\beta$                      &  114.35$\pm$1.33&170.65$\pm$2.78& 134.53$\pm$6.53& 31.00$\pm$1.57&104.89$\pm$1.90&  25.93$\pm$1.38& 108.09$\pm$2.85& 138.13$\pm$3.40&  26.63$\pm$1.44&  23.73$\pm$1.17 \\
{[\ion{He}{I}]$\lambda$4923}  &             --- &           --- &            --- &           --- &           --- &            --- &            --- &            --- &            --- &            ---  \\
{[\ion{O}{III}]$\lambda$4959} &   91.38$\pm$1.25&166.94$\pm$3.25& 148.81$\pm$6.01& 48.79$\pm$1.46&123.83$\pm$2.14&  44.44$\pm$1.51& 147.14$\pm$3.62&  81.76$\pm$2.70&  45.10$\pm$1.90&  35.11$\pm$1.20 \\
{[\ion{O}{III}]$\lambda$5007} &  255.55$\pm$1.84&507.77$\pm$5.48& 427.48$\pm$8.75&133.82$\pm$2.12&366.54$\pm$3.40& 122.08$\pm$2.10& 418.97$\pm$5.74& 232.02$\pm$4.09& 125.46$\pm$2.69& 103.15$\pm$2.94 \\
{[\ion{He}{I}]$\lambda$5017}  &    2.17$\pm$0.56&           --- &            --- &           --- &  2.84$\pm$0.90&            --- &            --- &            --- &            --- &            ---  \\
{[\ion{N}{I}]$\lambda$5199}   &             --- &           --- &            --- &           --- &           --- &            --- &            --- &            --- &            --- &            ---  \\
{[\ion{N}{I}]$\lambda$5201}   &             --- &           --- &            --- &           --- &           --- &            --- &            --- &            --- &            --- &            ---  \\
{[\ion{N}{II}]$\lambda$5755}  &             --- &           --- &            --- &           --- &           --- &            --- &            --- &            --- &            --- &            ---  \\
{[\ion{He}{I}]$\lambda$5875}  &   13.75$\pm$0.69& 21.58$\pm$0.99&  12.76$\pm$2.53&  4.75$\pm$1.31& 15.21$\pm$0.69&            --- &  11.88$\pm$1.17&  13.49$\pm$1.25&            --- &            ---  \\
{[\ion{O}{I}]$\lambda$6300}   &    2.13$\pm$0.36&  2.45$\pm$0.67&   9.40$\pm$2.15&           --- &           --- &            --- &            --- &   2.94$\pm$0.96&            --- &            ---  \\
{[\ion{S}{III}]$\lambda$6312} &    1.53$\pm$0.35&           --- &            --- &           --- &           --- &            --- &            --- &            --- &            --- &            ---  \\
{[\ion{N}{II}]$\lambda$6548}  &             --- &           --- &            --- &           --- &           --- &            --- &            --- &            --- &            --- &            ---  \\
H$\alpha$                     &  340.29$\pm$1.88&501.65$\pm$4.78& 278.06$\pm$5.87& 87.79$\pm$1.50&322.69$\pm$3.23&  76.11$\pm$1.15& 319.44$\pm$3.51& 398.01$\pm$4.49&  82.27$\pm$1.85&  77.05$\pm$3.80 \\
{[\ion{N}{II}]$\lambda$6584}  &    3.53$\pm$0.52&           --- &  14.07$\pm$2.00&           --- &           --- &            --- &   3.56$\pm$0.85&            --- &            --- &            ---  \\
{[\ion{He}{I}]$\lambda$6680}  &    3.91$\pm$0.36&  5.49$\pm$0.49&            --- &           --- &  3.88$\pm$0.54&            --- &   2.92$\pm$0.85&   4.50$\pm$0.75&            --- &            ---  \\
{[\ion{S}{II}]$\lambda$6717}  &    9.44$\pm$0.40&  7.27$\pm$0.51&  34.28$\pm$2.28&           --- &  1.66$\pm$0.55&   3.78$\pm$0.52&  11.69$\pm$0.95&  12.62$\pm$1.12&            --- &            ---  \\
{[\ion{S}{II}]$\lambda$6731}  &    6.71$\pm$0.32&  4.81$\pm$0.45&  24.86$\pm$1.94&           --- &  1.94$\pm$0.48&            --- &   8.28$\pm$0.84&   7.03$\pm$0.82&   2.38$\pm$0.47&            ---  \\
{[\ion{He}{I}]$\lambda$7067}  &    2.31$\pm$0.27& 11.90$\pm$0.67&            --- &           --- & 12.85$\pm$1.04&            --- &   3.27$\pm$0.81&            --- &            --- &            ---  \\
{[\ion{Ar}{III}]$\lambda$7136}&    4.34$\pm$0.33&  2.70$\pm$0.65&   6.30$\pm$1.66&           --- &  3.11$\pm$0.62&   0.98$\pm$0.32&   7.67$\pm$1.30&   4.64$\pm$1.13&            --- &            ---  \\
{[\ion{O}{II}]$\lambda$7320}  &             --- &           --- &   3.65$\pm$1.17&           --- &           --- &            --- &   1.53$\pm$0.48&   3.17$\pm$0.92&            --- &            ---  \\
{[\ion{O}{II}]$\lambda$7330}  &             --- &           --- &            --- &           --- &           --- &            --- &            --- &            --- &            --- &            ---  \\
{[\ion{Ar}{III}]$\lambda$7751}&             --- &           --- &            --- &           --- &           --- &            --- &            --- &            --- &            --- &            ---  \\
{[\ion{S}{III}]$\lambda$9068} &    9.02$\pm$0.29&           --- &  24.64$\pm$4.25&           --- &           --- &            --- &  10.14$\pm$0.82&   8.58$\pm$1.05&            --- &            ---  \\
{[\ion{S}{III}]$\lambda$9530} &   24.90$\pm$0.53&           --- &            --- &           --- &           --- &            --- &            --- &  20.72$\pm$1.31&            --- &            ---  \\
Pa8                           &    5.13$\pm$0.32&           --- &            --- &           --- &           --- &            --- &            --- &            --- &            --- &            ---  \\
\hline        
C(H$\beta$)                   &   0.098$\pm$0.017&0.083$\pm$0.024&0.000$\pm$0.066&0.037$\pm$0.068&0.140$\pm$0.026&0.081$\pm$0.069&0.090$\pm$0.035 &0.058$\pm$0.033 &0.145$\pm$0.075 & 0.207$\pm$0.087  \\
EW(H$\beta$)                  &    82.9$\pm$1.0 &  62.2$\pm$1.0 &   41.1$\pm$2.0 &  51.6$\pm$2.6 & 304.7$\pm$5.6 &   40.2$\pm$2.1 &   86.7$\pm$2.3 &   90.0$\pm$2.2 &   74.9$\pm$4.0 & 104.4$\pm$5.2   \\

\hline
\end{tabular}
\end{table}
\end{landscape}

\begin{figure*}
\resizebox{1.0\hsize}{!}{\includegraphics[angle=000]{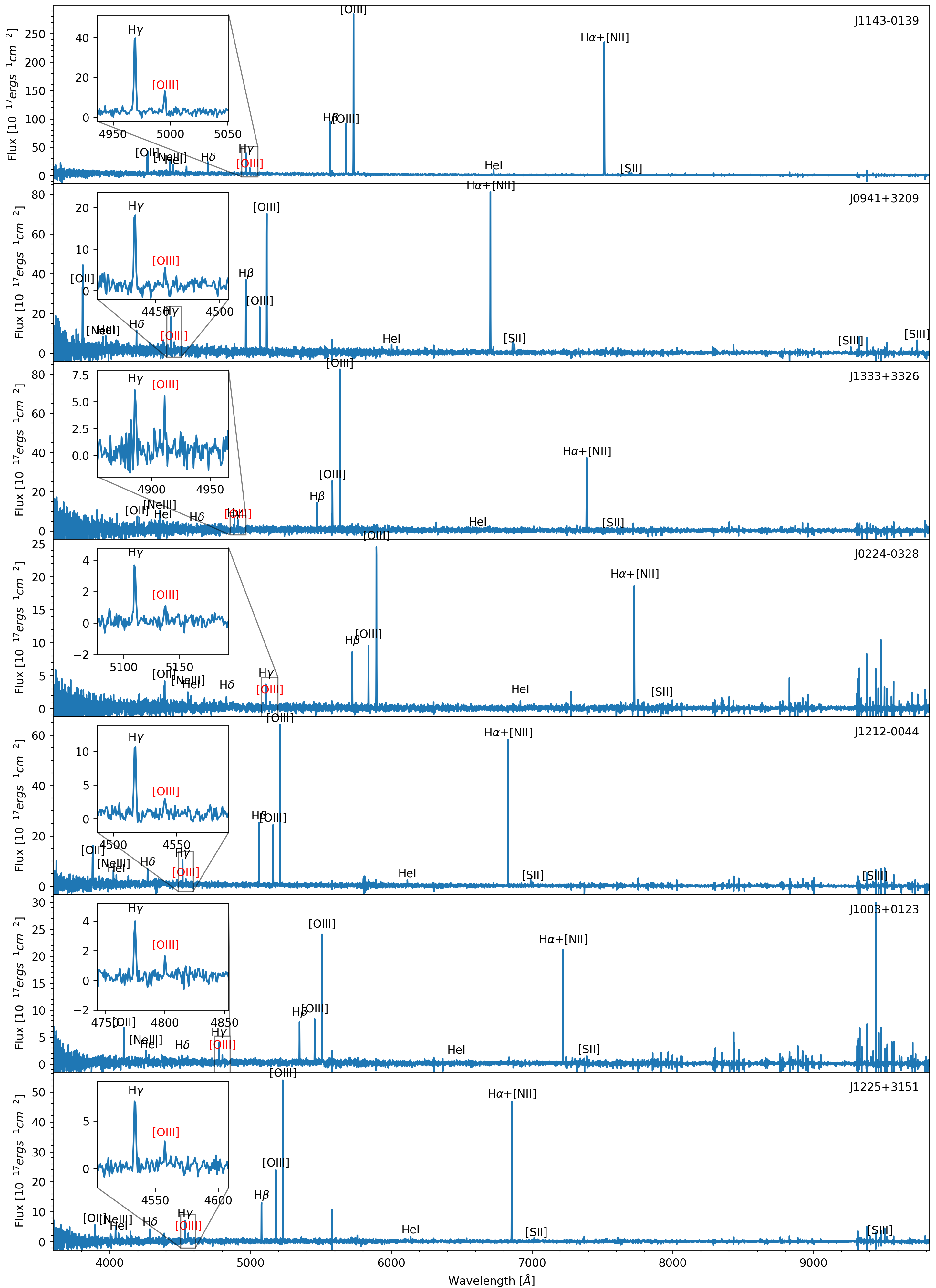}}
\caption{%
DESI spectra of our sample of galaxies. Selected emission lines are labelled.
}
\label{figure:spectra1}
\end{figure*}

\begin{figure*}
\resizebox{1.0\hsize}{!}{\includegraphics[angle=000]{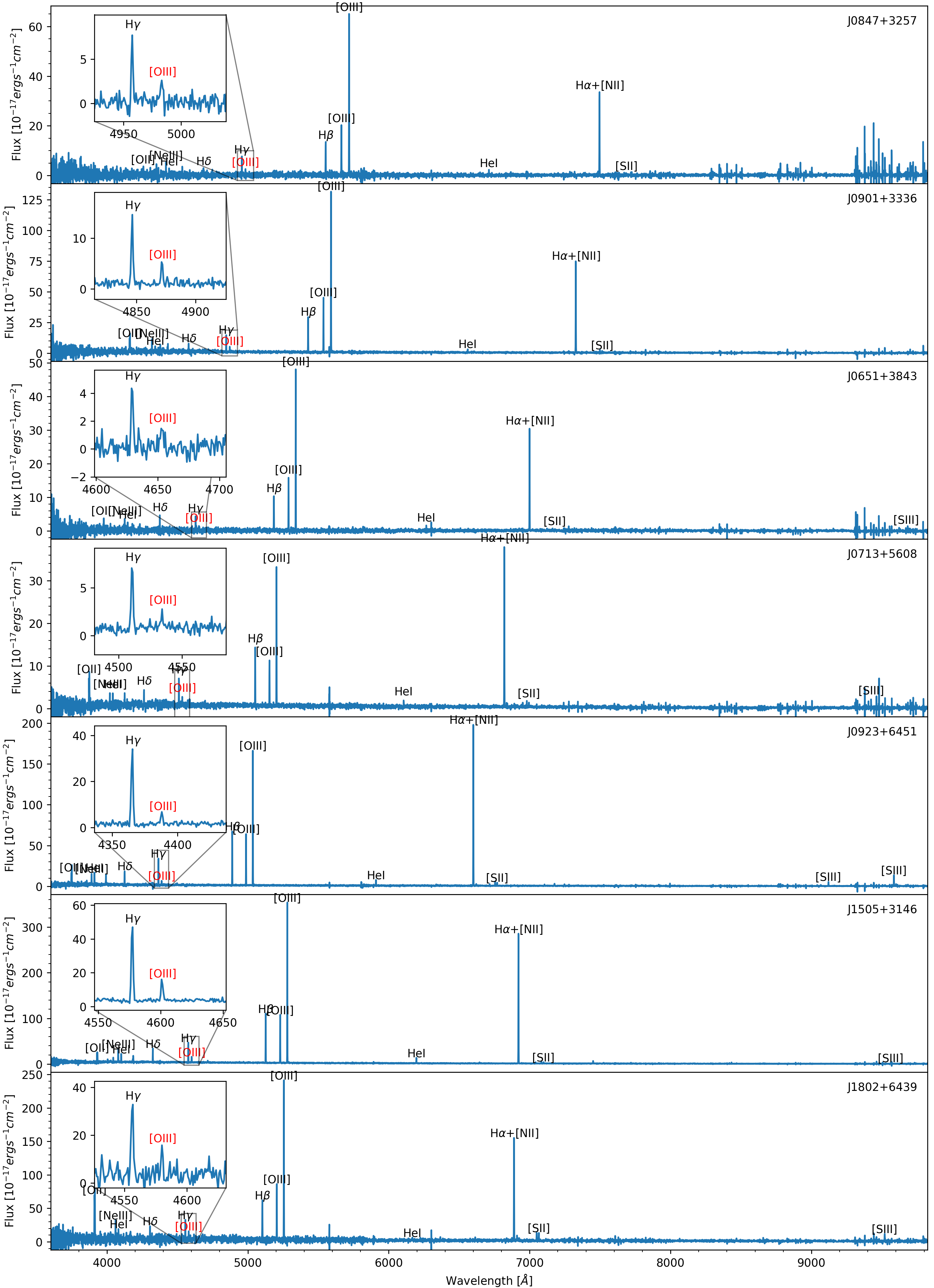}}
\caption{%
DESI spectra of our sample of galaxies. Selected emission lines are labelled (continued).
}
\label{figure:spectra2}
\end{figure*}

\begin{figure*}
\resizebox{1.0\hsize}{!}{\includegraphics[angle=000]{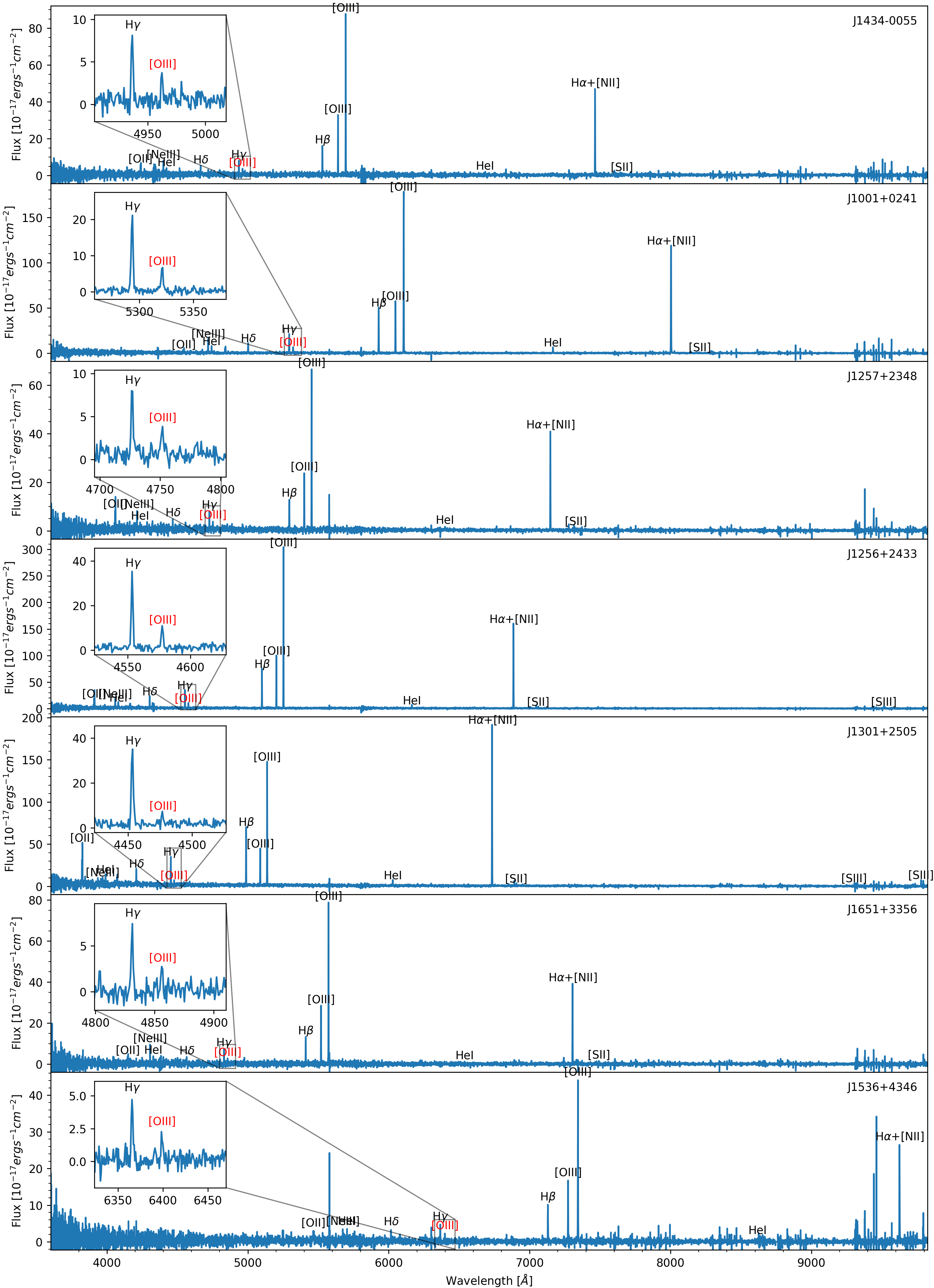}}
\caption{%
DESI spectra of our sample of galaxies. Selected emission lines are labelled (continued).
}
\label{figure:spectra3}
\end{figure*}

\end{appendix}
 
\end{document}